# Extreme thunderstorm ground enhancements registered on Aragats in 2023

*A. Chilingarian, B. Sargsyan, T. Karapetyan, D. Aslanyan, S. Chilingaryan, L. Kozliner, Y.Khanikyanc*

A. Alikhanyan National Lab (Yerevan Physics Institute), Yerevan 0036, Armenia

**Abstract**

In 2023, a series of intense Thunderstorm Ground Enhancements (TGEs) were recorded on Mount Aragats in Armenia, with five events exceeding the fair-weather cosmic ray flux by more than 100%. This study comprehensively analyzes these TGEs, investigating the atmospheric conditions and electric fields contributing to their occurrence. Key insights include discovering relationships between TGEs and atmospheric electric fields, recovering electron and gamma-ray energy spectra, and the impact of nearby lightning activity. The findings offer a deeper understanding of TGEs' role in atmospheric physics and its synergy with high-energy astrophysics.

**Plain Language Summary**

In 2023, our research team on Mount Aragats in Armenia detected a series of dramatic surges in particle fluxes linked to thunderstorms, with intensities rising to ten times more than on a clear day. This boost is due to strong electric fields in the clouds that act like natural accelerators, speeding up electrons to create Relativistic Runaway Electron Avalanches (RREAs). When/if they reach the Earth's surface, these avalanches manifest themselves as Thunderstorm Ground Enhancements (TGEs), large peaks in the time series of electrons, gamma rays, and neutrons registered by particle detectors.

Our findings revealed a clear pattern of TGEs initiated by RREAs, shedding light on the conditions for their formation. Surprisingly, thunderstorm-related electric fields can be as strong as ≈2.0 kV/cm, even at heights as low as below 50 meters above the ground. This observation and the consistent power of these natural electric field accelerators during thunderstorms unveil a new level of structured behavior in atmospheric electric fields that we hadn't recognized before.

1. Introduction
2. 

In 2023, 56 intense Thunderstorm Ground Enhancements (TGEs) (Chilingarian et al., 2010; 2011) were observed on Mount Aragats, Armenia. The largest of these is marked by dramatic tenfold increases in cosmic ray fluxes compared to fair-weather conditions was observed on May 23. This study delves into the atmospheric processes driving these phenomena, shedding light on the generation of high-energy particles during thunderstorms.

TGEs are initiated when the electric field within thunderclouds surpasses a critical threshold, accelerating free atmospheric electrons. This process, known as Relativistic Runaway Electron Avalanches (RREA, Gurevich et al., 1992), results in the multiplication of energetic electrons, producing bremsstrahlung gamma rays. These gamma rays can further interact through photonuclear reactions to generate neutrons (Chilingarian et al., 2012), contributing to the complex mixture of particles observed during a TGE event.

The dynamic structure of thunderclouds plays a pivotal role in TGE development (Dwyer, 2003; Babich et al., 2004). Specifically, the formation of charged layers within the clouds - such as the main negatively charged layer (MN) and its corresponding induced mirror charge in the Earth (MIRR1), along with an additional positively charged region (LPCR) and its mirror (LPCR-MIRR2) - creates electron and positron accelerated (decelerated) dipoles. These dipoles undergo continuous changes, changing the modes of particle acceleration as the cloud's charge structure evolves.

Over the past decade, Mount Aragats and other sites like Mount Lomnicky Stit, Zugspitze, and Mount Musala have documented nearly a thousand TGE events, with some showing cosmic ray flux enhancements up to 100 times the background levels (Chum et al., 2020).

However, till now, only TGEs observed by the Aragats Space Environmental Center (ASEC, Chilingarian et al., 2005) have entered the databases accompanied by measurements done by field meters, weather stations, and all-sky cameras. The intensities and energy spectra of TGE electrons, gamma rays, and neutrons are measured by approximately 100 channels of particle detectors. This study utilizes extensive datasets from Aragats particle detectors, near-surface electric field and weather parameter measurements, all-sky camera shots, and lightning location data. Analysis of hundreds of TGEs registered on Aragats provides valuable insights into the interplay between the near-surface electric fields, geomagnetic field variations, lightning activity, and the weather conditions on TGE occurrences (Chilingarian et al., 2022a).

We categorize TGEs based on their characteristics—the flux enhancement relative to fair-weather values, duration, electron and gamma-ray energy spectra, presence of lightning flashes abruptly terminating TGEs, stability of the flux, and unique TGEs coincide with observable optical glows registered by all-sky cameras (Chilingarian et al., 2022b). One of the most important tasks is understanding the electric field's vertical and horizontal locations supporting large TGE development at remote sites.

The precise experimental efforts on Mount Aragats are integral to uncovering the enigmatic high-energy physics within atmospheric plasmas (HEPA, Dwyer et al., 2012). Comprehending these phenomena enhances our understanding of natural particle acceleration mechanisms within Earth's atmosphere and space plasmas, thus offering broader implications for atmospheric science and astrophysical research.

## 3. Continuous Monitoring of Electric Fields and Particle Fluxes on Mount Aragats

In 2023, 56 TGEs (see the raw data in Chilingarian et al., 2024a) were registered by particle detectors, largely enhancing the 11-year mean of 35.5. The lowest number of registered TGEs was in 2019 – 15. The count rate of 5 TGEs exceeds 100% above the fair-weather value measured by the STAND1 detector (Chilingarian et al., 2022c)—the intensity of ever largest TGE of 23 May overpass fair-weather intensity ten-fold. The fluence of TGE on 27 May reaches 3.8 particles/cm$^2$. The largest TGEs sustain a very stable flux for a few minutes, demonstrating the astonishing stability of atmospheric electron accelerators.

Figure 1 shows 1-minute time series measurements of particle fluxes, near-surface electric field (NSEF), and distances to lightning flashes over a year. Particle fluxes obtained by the STAND3 detector comprise four identical vertically stacked plastic scintillators, each with an area of 1 m² and a thickness of 3 cm. By measuring different coincidences of the scintillators, we can select TGE particles with different energies. The coincidence "1000" isolates electrons and gamma rays with energies above 5 MeV, while, for instance", "1110" targets electrons above 30 MeV with very small gamma-ray contamination. Electrons with lower energies are absorbed within the detector's body.

Intense thunderstorms on Mount Aragats, as seen in Fig. 1, primarily concentrated in May-July, when the NSEF reached and overpassed +/- 30 kV/m. NSEF in fine weather is ≈0.15 kV/m. In winter months, as expected, thunderstorms are much rarer. The weather conditions supporting the emergence of large TGEs and relations between NSEF and TGE are far from well understood (Williams, 2023). However, due to the current limitations in directly measuring atmospheric electric field (AEF) above Aragats, we utilize NSEF measurements represented by the time series in Fig. 1 as its proxy (black curve).

The 24/7 monitoring of NSEF by the network of BOLTEK's EFM 100 electric mills (BOLTEK, 2024) on Aragats offers crucial insights into rapid electrification changes in the lower atmosphere. The most significant advantages of NSEF monitoring include its ability to reveal the cloud charge structure in the lower atmosphere directly associated with emerging dipoles in thunderclouds (see for details Chilingarian et al., 2024b, 2024c). The blue curve in Fig. 1 depicts a 1-minute time series of count rate; TGEs are indicated by sharp lines above the background. Red lines in Fig.1 indicate nearby lightning flashes.

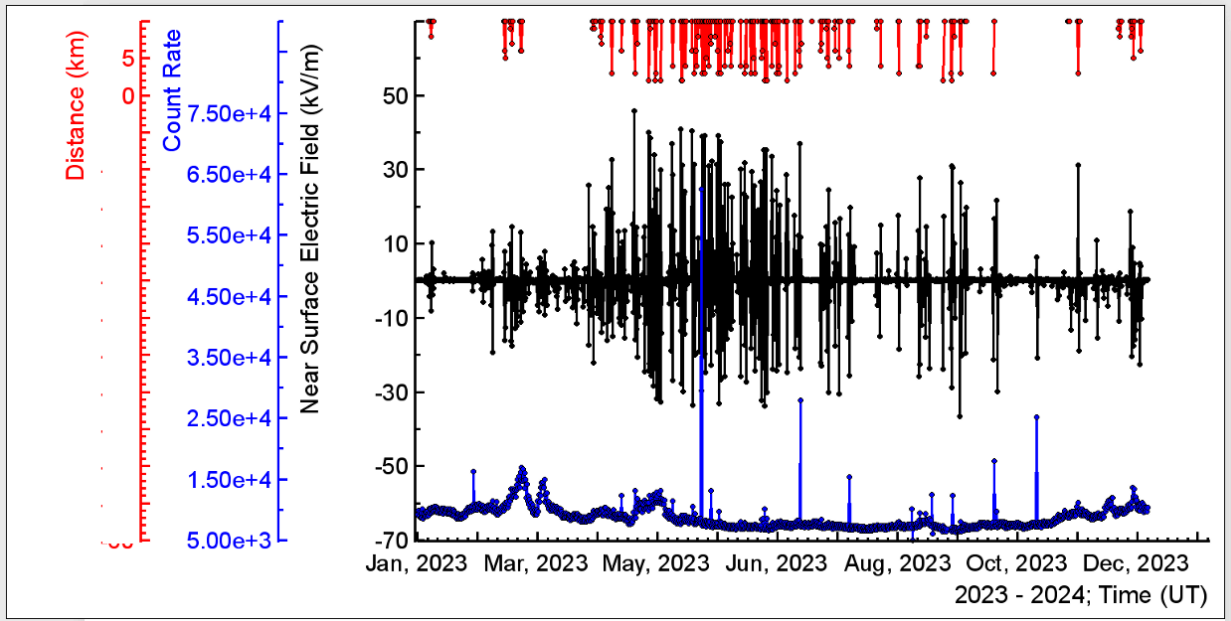

**Figure 1.** Black shows disturbances of the NSEF measured by EFM 100 electric mills produced by the BOLTEK firm; blue shows a time series of 1-minute count rates of a STAND3 plastic scintillator with a 1 m² area and 3 cm thickness (1000 coincidence, signal only in the upper scintillator); red shows distances to lightning flashes.

Five TGE events surpass the limit of 100% enhancement (by STAND3's "1000" coincidence), shown above the red line in Fig.2. The huge TGE of 23 May (≈800% enhancement) is the largest among ≈ 700 TGEs during the HEPA research started in 2008 on Aragats.

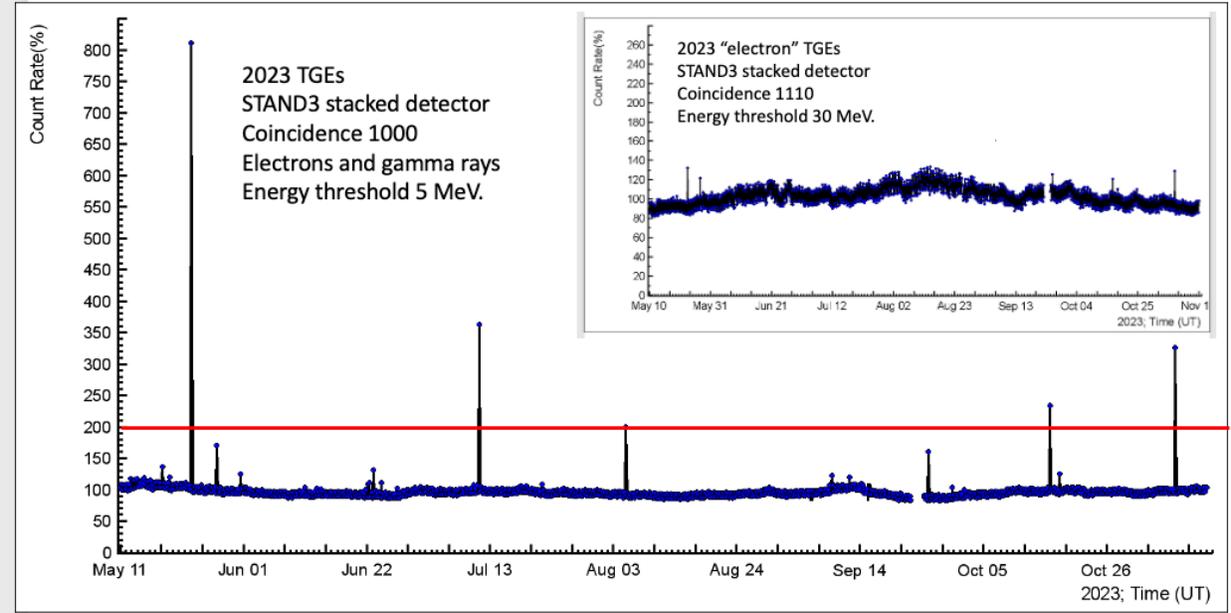

**Figure 2.** TGE events occurred on Aragats from May to November 2023—1-minute time series of the STAND1 detector. The "1000" coincidence selects low-energy electrons and gamma rays, and the "1110" coincidence (in the inset) - high-energy electrons.

The largest "electron TGEs" selected by the "1110" coincidence of the STAND3 detector (targeting by electrons with energies above 30 MeV) occurred on 23 May and 6 November, see the inset to Fig.2. Near 700 additional high-energy electrons were registered at the minute of maximum flux at both dates. The relations of particle fluxes, electric fields, and atmospheric conditions are not straightforward. A large diversity exists in particle fluxes, NSEF strengths, and atmospheric conditions during TGEs. You cannot find 2 TGEs with the same parameters. It makes revealing causal relations among measured parameters rather tough. However, multivariate correlation analysis helps to find useful relations between measurements.

## 4. Detailed description of the largest TGEs

In Figure 3, we depict the network of STAND1 particle detectors placed at the vertices of a triangle with lengths of 137 m, 226 m, and 240 m. The STAND1 detector consists of three layers of 1 cm thick, one m² sensitive-area scintillators stacked vertically. Additionally, one 3 cm thick plastic scintillator of the same type stands apart. The scintillator light is re-radiated into the long-wavelength region of the spectrum by the spectrum-shifter fibers and transmitted to the photomultiplier (PMT, FEU 115 M). The STAND1 detector is adjusted by changing the high voltage applied to the PMT and setting the shaper-discriminator thresholds. The discrimination level is chosen to ensure high signal detection efficiency and maximum suppression of photomultiplier noise. Based on simulations and calibration experiments, we estimate the efficiency of the STAND1 upper scintillator presented in the paper for charged particles to be approximately 95% with energy thresholds of around 1 MeV.

The fast, synchronized data acquisition (FSDAQ, Pokhsraryan, 2016) system allows for registering particle fluxes and near-surface electric field disturbances. National Instrument's MyRio board (Chilingarian et al., 2024d) offers eight analog inputs, four analog outputs, thirty-two digital I/O lines, FPGA, dual-core ARM Cortex-A9 processor, and GPS. We can perform high-speed signal processing, control, inline signal processing, and custom timing and triggering using reconfigurable FPGA technology. We use eight digital inputs from three MyRio boards to feed signals from the STAND1 network (four channels for each board) and EFM 100 electric mill (by WiFi). Each MyRio board generates an output signal that includes the 50 ms count rates from four scintillators, near-surface electric field value, and the GPS timestamp of the trigger signal. This way, the count rates and NSEF strengths measured by STAND1 and EFM 100 networks are synchronized on a millisecond time scale. The ADEI data analysis platform allows for multivariate visualization and correlation analysis of all-time series collected during the 12 years of STAND1 network operation (Chilingarian et al., 2008).

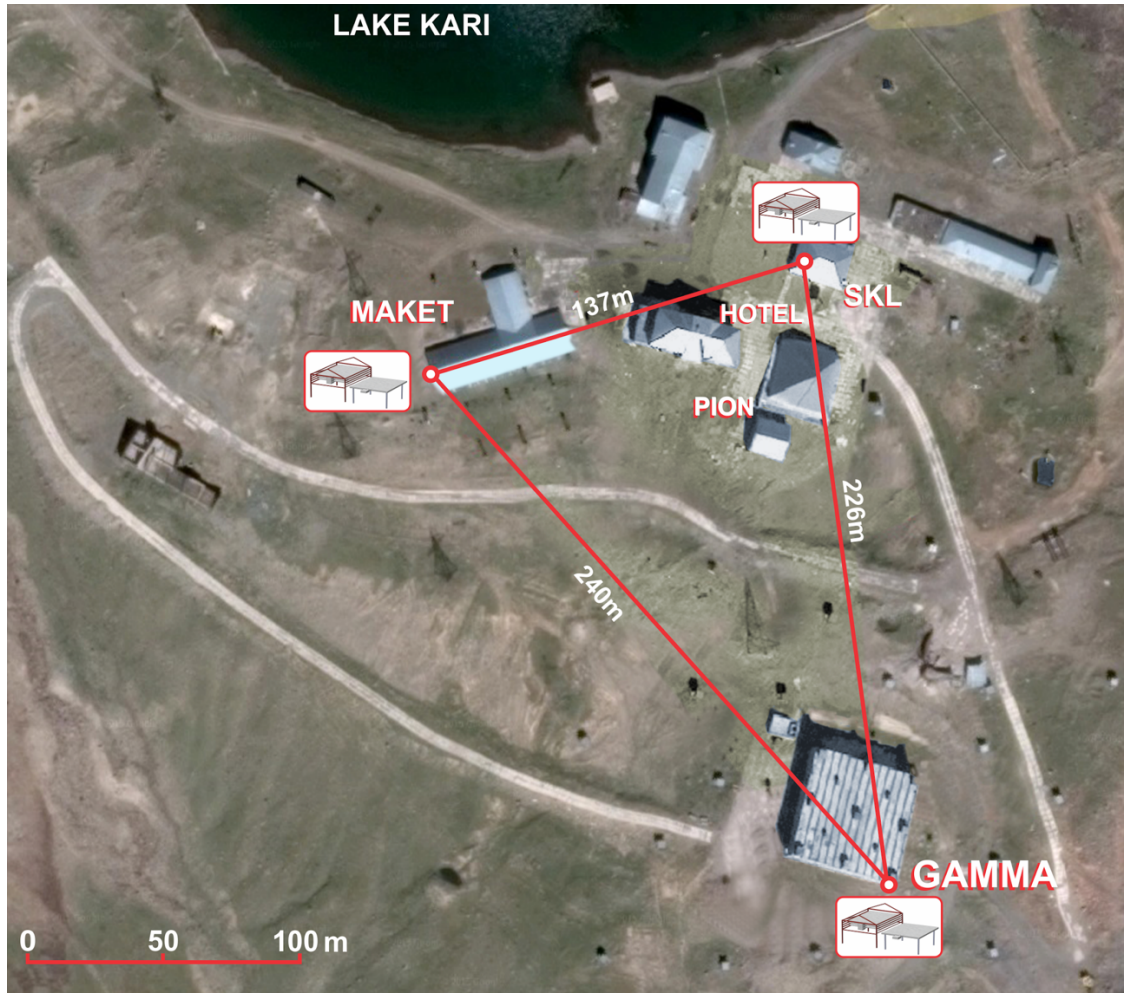

**Figure 3. STAND1 particle detector network at Aragats research station, 3200 m above sea level.**

In Figure 4, we present the four largest Summer and Autumn TGEs registered by the upper scintillators of the STAND1 network depicted in Fig. 3. There is a difference in the count rate rise and decay in Summer (Figs 4a and 4b) and Autumn TGEs (Figs 4c and 4d). Summer TGEs are highly uniform; in the Autumn TGEs, we see asymmetry between detectors located at a distance ≈of 100 m at the highland near Kari Lake (red and blue curves) and remote detector located ≈250 m lower on the slope of the mountain opened to the Ararat Valley. Previously, we detected asymmetry of the near-surface electric field (NSEF) measured at these locations by BOLTEK's EFM 100 sensors (Chilingarian et al., 2022e) network.

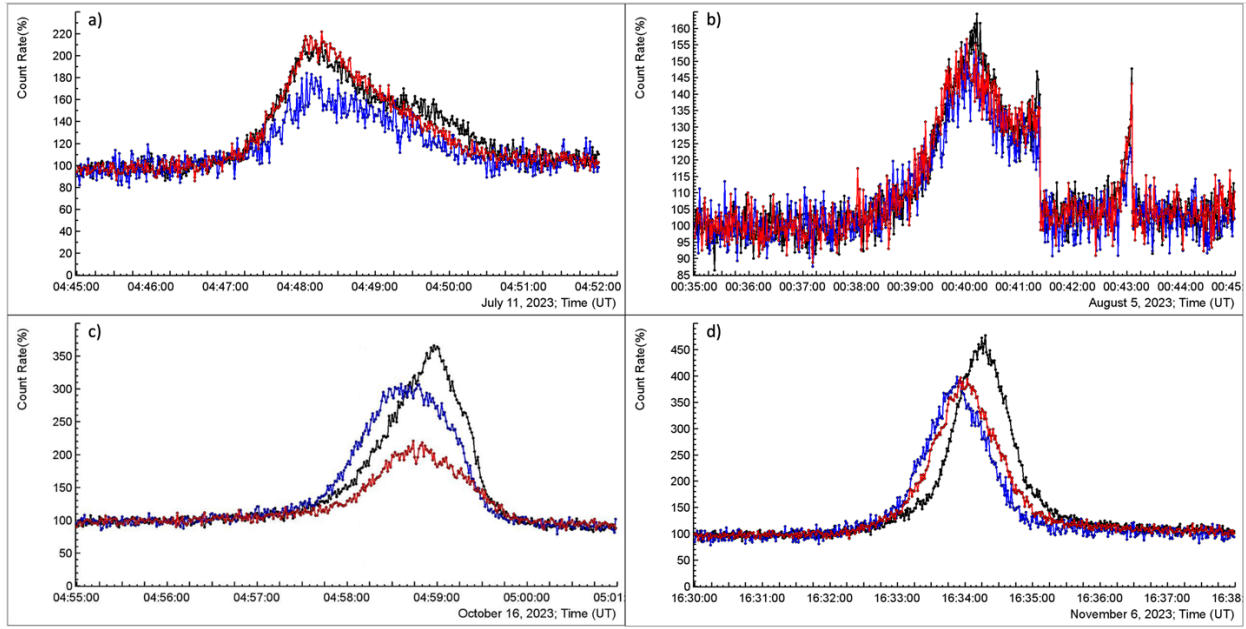

**Figure 4.** One-second time series of count rates of 1 cm thick and 1 m² area upper scintillators of the STAND1 distributed particle detector network. Black – STAND1 unit on the roof of GAMMA experimental hall; blue – nearby MAKET, and red nearby SKL experimental halls.

Despite the variations in the enhancements of count rates shown in Figure 4, the particle flux appears consistent across an area of approximately 50,000 square meters covered by the STAND1 network. In previous studies (Chilingarian et al., 2022c), we reported TGEs recorded at the Aragats and Nor Amberd research stations within several minutes, which are 13 kilometers apart. This suggests that a large mesoscale electric field is overlying both stations, extending over several cubic kilometers in the atmosphere, resulting in particle fluxes on Earth's surface that cover many square kilometers. The location of the detectors can explain the difference in the percentage of flux enhancement. STAND1, situated on the roof of GAMMA, is exposed from all sides, whereas the other two detectors are located near buildings and are partially obstructed by walls.

Figure 5 shows the count rate enhancement of the four largest TGEs (blue curve), along with disturbances of NSEF (black curve) and distances to the lightning flash (red lines). Count rates are measured by 1 m² area 1 cm thick scintillators of STAND1 network located on the roof of the GAMMA experimental hall. The duration of the largest TGEs is 3-5 minutes; 2 TGEs were abruptly terminated by the lightning flash at maximum particle flux (Fig. 5a, one of them occurred on August 5, is not shown); 3 TGEs occurred when NSEF was in the negative domain, Summer TGEs (11 July and 5 August) – in the positive domain. Lightning flashes occurred before and after TGE. During all five largest TGEs, no lightning flashes were observed within a 15 km radius. See our classification of the lightning types terminating TGE (Chilingarian et al., 2024e).

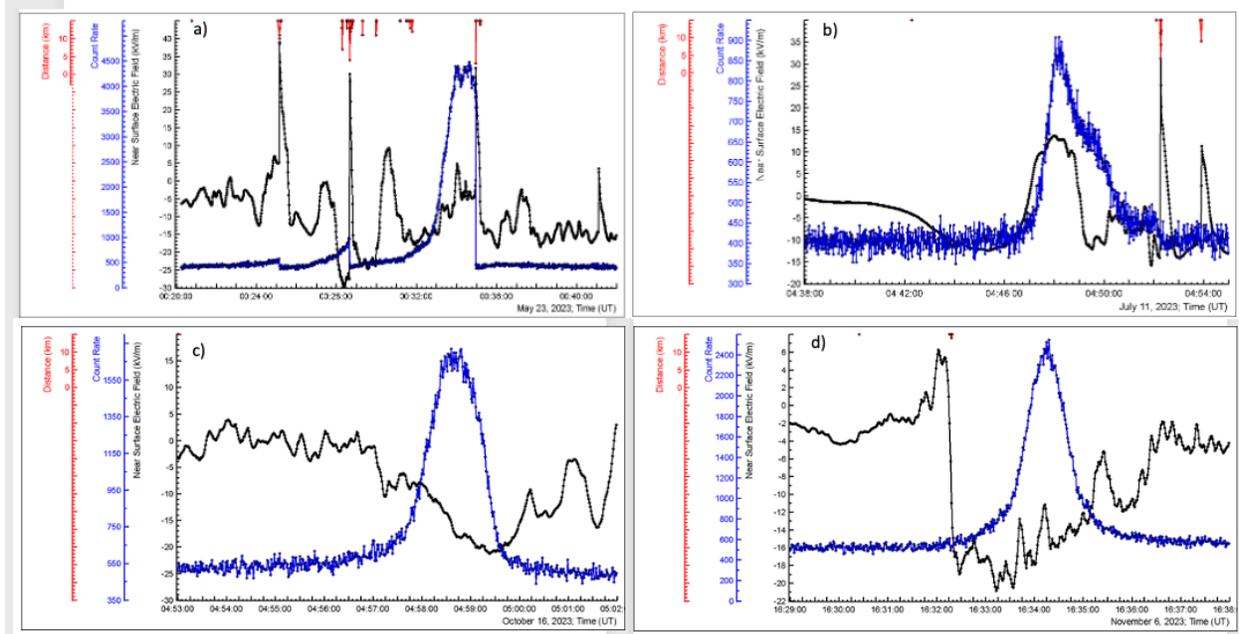

**Figure 5. Largest TGEs of 2023. Black – disturbances of NSEF. Blue – 1-minute time series of 1 cm thick, 1 m$^2$ scintillator of the STND3 detector. "1000" coincidence – signal only in the upper scintillator of 4 identical stacked vertically sensors—red – distance to the lightning flash.**

In Figure 6, we present four 50 ms time series measured by the stacked STAND1 network's scintillator (corresponding to Fig. 5). The network covered a 50,000 m$^2$ area over the Aragats research station, and the TGE parameters measured by three detector setups are very close. The count rates of STAND1 network detectors are higher than those of the STAND3 detector located in the SKL experimental hall. The energy threshold of the 1-cm thick outdoor detector is ≈1 MeV, and the 3-cm thick scintillator in the building is ≈ 5 MeV.

There are distinctions in the NSEF-TGE relation in different seasons. The spring TGE (Fig. 5a) started just after two nearby lightning flashes at small positive NSEF, which quickly turned to negative. TGE flux rises with enhancing negative NSEF till abruptly terminated by lightning flash. The TGE in summer (Fig. 5b) occurred during a positive NSEF, first decreasing smoothly to the negative domain and then quickly turning positive at the TGE flux rise. After being in the positive domain for a few minutes, TGE turns to the negative domain and stays there until TGE smoothly ends. Afterward, in 5 minutes, two nearby flashes were detected. Minimal lightning activity was present during Autumn TGEs, which displayed symmetric and smooth bell-like shapes. (See Figs 5c and 5d). Both TGEs occurred during negative NSEF; however, during the first one (Fig 5c), NSEF decreased during TGE, and during the second (Fig. 5d) increased. Thus, despite the shapes and durations of TGEs being close, the relation to NSEF (a proxy of the AEF) is very different. Thus, cloud accelerators operate in different modes, accelerating electrons and decelerating positrons, and vice versa, accelerating positrons and decelerating electrons depending on emerging charge structures in the lower atmosphere (Chilingarian et al., 2024b; 2024c).

To gain further insight into TGEs, we investigate the stability of the particle flux at its maximum values. Figure 6 compares the 50 ms time series of TGE count rates at maximum flux with the time series measured at the same time the day before during fair weather. All 4 TGE time series are well above the fair-weather time series. In the legends, we present the selected intervals of TGE (from 30 s to 1 minute), count rate means, standard and relative errors, the enhancement in percent, and its significance in the number of standard deviations above the mean value measured at fair weather. We also present the equivalent significance for the 1-minute time series for comparative purposes.

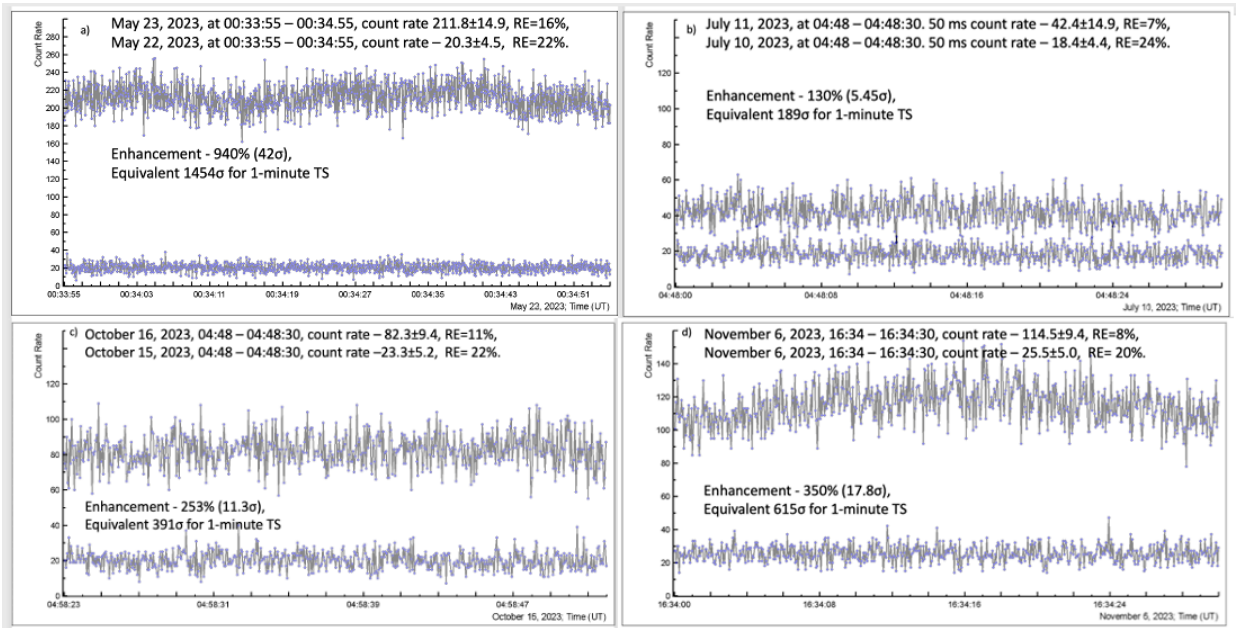

**Figure 6. 50 ms time series of count rates of the STAND1 detector's upper scintillator, which is 1 cm thick and has a 1 m² area. The lower curve in each frame corresponds to the count rate measured in fair weather at the same time one day before TGE. The legend includes each TGE's mean count rates, standard errors, and significance.**

As shown in Figure 6 a-d, the TGE flux is very stable - the relative errors for all 4 TGEs are smaller than the ones calculated for the ambient cosmic ray flux (a day earlier). Thus, despite the fast-changing atmosphere environments, electron accelerators sustain stable particle flux over at least 50,000 m² (the area covered by the STAND1 network).

Figures 7a-7d present the integral energy spectra of the 4 out of 5 largest TGEs. We continuously elaborate particle spectrometers operating on Aragats, starting from estimating energy spectra by four energies, corresponding to the energy thresholds of used detectors (Cilingarian et al., 2013) until solving the inverse problem using detailed calculation of the detector response function with GEANT4 package (Agostinelli et al., 2003). The energy spectra were recovered from the measured energy release spectrum by the CUBE detector, detailed in (Chilingarian et al., 2024f). We use

two 20 cm thick and 0.25 m² area plastic scintillators for spectrometry and a 1 cm thick and 1 m² area scintillator fully covering spectrometric scintillators for vetoing charged and fluxes. Using coincidences of 2 scintillator operations, we isolate charged and neutral fluxes. The methodology of spectra recovery from the energy release histograms was the same as for the ASNT spectrometer (Chilingarian et al., 2022f) and is described in (Chilingarian et al., 2024f) in detail. The Logarithmic Amplitude-Digit Converter (LADC), used in CUBE electronics, has a unique feature that allows to measure of energy deposits in a wide dynamic range of input signal amplitudes (corresponding to particle energies up to 100 MeV)

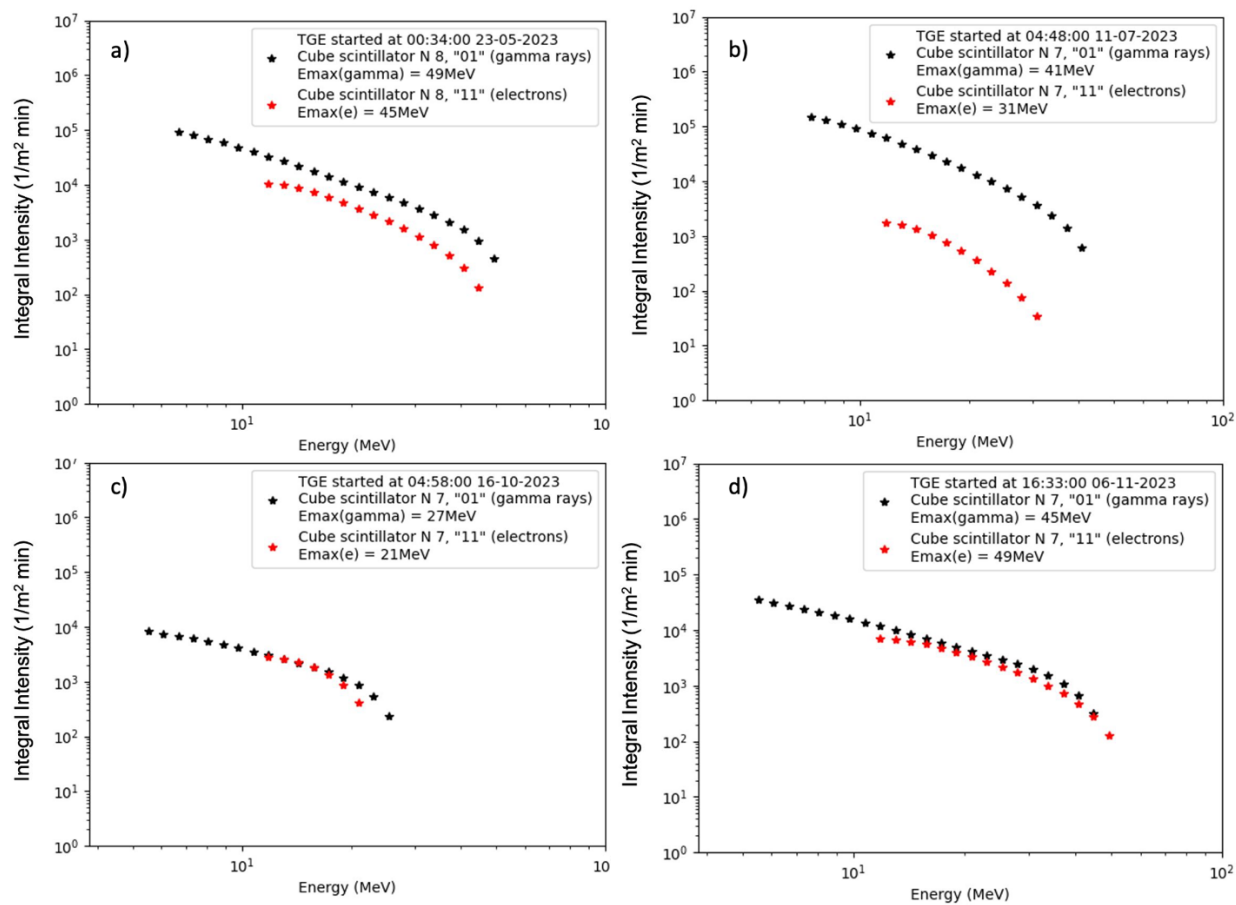

**Figure 7. The integral energy spectra of the largest TGEs of 2023**

The electron and energy spectra of Autumn TGEs (Figs 7c and 7d) are very close in contrast with Spring-Summer TGEs, which differ in order of magnitude (Figs 7a and 7 b). This hint can help to estimate the extension of the electric field above the detectors, which will be considered in detail in the next section.

## 4. Free passage distance and distance to the cloud base

We have calculated several parameters from the initial measurements to estimate atmospheric electric field strength and location. These parameters are the distance to the cloud base and free passage distance (FPD) from the ground where the RREA avalanche leaves the accelerating intracloud electric field. The distance at which the strong accelerating field is terminated, FPD, is calculated by an empirical equation (1) confirmed by simulations (Chilingarian et al., 2021a).

FPD (meters) = $(C_1 * E^\gamma_{max} – E^e_{max})/C_2$   (1),

Coefficients $C_1$ and $C_2$ are estimated to be 1.2 and 0.2, respectively. TGE simulations suggest that the maximum energy of electrons going out of the electric field is 20% higher than that of gamma rays (Chilingarian et al., 2021a). Therefore, we can estimate the maximum energy of electrons leaving the field by $C_1 * E^\gamma$ max. Furthermore, we assume that the maximum energy of gamma rays does not change significantly when they travel 100 m or less in the atmosphere. Also, we assume that electrons lose approximately 0.2 MeV per m at altitudes of about 3000m. We conducted multiple simulations of electron-gamma ray avalanches to verify the accuracy of equation (1) and detect any potential methodological errors. We store the particle energies and solve the inverse problem to recover the RREA characteristics from the measured TGE. We utilize CORSIKA simulations (Heck et al., 1998) with varying electric field strengths and termination heights to achieve this. Subsequently, we apply all experimental procedures to the obtained samples to estimate the maximum energies of electrons and gamma rays (Chilingarian et al., 2021a, supplemented materials). Once we have calculated the FPD parameter, we compare it to the "true" value in the simulation. Based on this comparison, we estimate the method's standard error to be 50 meters - a conservative estimate. The cloud base height is recovered by calculating the air temperature and dew point spread according to the well-known approximate equation (2) (CLOUD BASE, 2024). The difference (or spread) between the air temperature and the dew point indicates how much cooling is needed for condensation. This method assumes a linear and uniform decrease in temperature with altitude, which might not always be the case in real atmospheric conditions with local variations.

H(m) ≈ (Air temperature at surface {°C} − dew point temperature {°C}) × 122   (2)

Table 1 displays TGE significances in percent relative to fair weather flux, TGE duration, weather parameters, and locations of the strong accelerating electric field within the thundercloud. The third to fifth columns show the percentage of TGE enhancement of STAND3's "1000" coincidence and absolute count rate enhancement of coincidences "1100" and "1110" targeting high-energy electrons; the sixth column shows the percentage of enhancement of SEVAN detectors "100" coincidence; seventh column – TGE duration by the STAND3's "1000" coincidence; eighth column – outside temperature. The ninth and tenth columns contain estimated parameters of the intracloud electric field, including cloud base height and free passage distance, estimated by equations (1) and (2). The last column shows the distance to the nearest lightning flash during TGE. By the bold fonts, we denote the five largest TGEs; those count rates exceed 100% of fair weather.

**Table 1. Detailed information on the 14 most significant TGEs of 2023; count rate significance was calculated by STAND3 1000 coincidence.**

| Date | Time (UT) | STAND3 Coins1000 % | Coins. 1100 count | Coins. 1110 count | SEVAN Coins. 100 % | Dur. min | Temp C° | Cloud base m | FPD m | EFM km |
|---|---|---|---|---|---|---|---|---|---|---|
| 2023-04-13 | 02:08 | 36.0 | 867 | 450 | 6.0 | 7 | -4.7 | 61.0 | | 17.0 |
| 2023-04-19 | 15:27 | 48.6 | 1200 | 432 | 7.0 | 8 | -1.8 | 48.8 | | 3.0 |
| 2023-05-08 | 02:02 | 28.0 | 729 | 287 | 3.9 | 26 | -1.9 | 73.2 | | 17.0 |
| 2023-05-18 | 05:19 | 25.8 | 553 | 240 | 6.4 | 8 | 1.9 | 231.8 | | 7.0 |
| **2023-05-23** | **00:31** | **675.0** | **4020** | **1080** | **69.7** | **5** | **0.1** | **36.6** | **70** | **3.0** |
| **2023-05-27** | **15:10** | **74.9** | **1400** | **650** | **15.5** | **33** | **0.5** | **73.2** | **57** | **14.0** |
| 2023-05-31 | 08:35 | 29.4 | 657 | 237 | 13.4 | 13 | 1.1 | 122.0 | | 4.0 |
| 2023-06-22 | 08:18 | 20.7 | 387 | 180 | 7.3 | 30 | 1.6 | 73.2 | | 16.0 |
| 2023-06-23 | 03:48 | 41.0 | 646 | 313 | 18.0 | 13 | 2.0 | 97.6 | | 20.0 |
| 2023-06-23 | 05:08 | 32.3 | 568 | 184 | 10.2 | 17 | 3.1 | 97.6 | | 12.0 |
| 2023-06-24 | 08:31 | 20.9 | 360 | 250 | 7.6 | 14 | 4.8 | 36.6 | | 2.0 |
| **2023-07-11** | **04:45** | **261.0** | **900** | **200** | **70.7** | **7** | **4.1** | **61.0** | **45** | **18.0** |
| **2023-08-05** | **00:38** | **115.0** | **660** | **0** | **34.4** | **6** | **7.7** | **170.8** | **215** | **6.0** |
| 2023-09-09 | 05:23 | 22.3 | 470 | 270 | 9.4 | 6 | 3.2 | 85.4 | | 14.0 |
| 2023-09-25 | 13:41 | 86.7 | 750 | 400 | 38.7 | 14 | 3.2 | 366.0 | | 22.0 |

| 2023-10-16 | 04:57:00 | 131.2 | 1530 | 500 | 25.2 | 10 | -0.6 | 61.0 | 57 | 3.0 |
| 2023-10-17 | 20:19 | 30.4 | 563 | 274 | 7.8 | 15 | -0.8 | 85.4 | | 18.0 |
| 2023-11-06 | 16:33 | 234.4 | 2500 | 800 | 37.7 | 10 | 1.1 | 97.6 | 25 | 14.0 |

Four of the five biggest TGEs occurred during the night or early morning, whereas only one occurred during the day. Additionally, two of the largest TGEs were terminated by a lightning flash. The TGEs during the summer happened during positive NSEF, whereas the rest happened during the negative. The number of electrons selected by coincidences of the STAND3 detector indicates the presence of electrons in TGE flux. On 23 May and 6 November, the number of particles selected by the "1110" coincidence was 1080 and 800, respectively, with corresponding maximum energies of TGE electrons being 59 and 49 MeV. Figure 8 presents the regression function of the free path distance (FPD) and the distance to the cloud base for the six largest TGEs. While it is expected that they should be correlated, the exact relationship shown in Fig. 8 provides the first direct evidence of it. Five out of the six largest TGEs occurred when both FPD and distance to the cloud base were within 100 m, and for all five of them, we could recover the energy spectrum of TGE electrons. Only the TGE with both parameters equal to approximately 200 m did not allow electron spectrum recovery, which was also expected.

We recovered TGE particle energy spectra above the roof of the building where the spectrometers are placed. The maximum energies of electrons and gamma rays for the biggest TGEs reach approximately 50 MeV, as indicated by the picture frames' logos. However, the maximum energies of electrons at the entrance from the accelerating field are higher due to ionization losses in the dense atmosphere. For the 5 out of 6 largest TGEs shown in Figure 8, FPD and cloud base height are around 50-60 meters. Therefore, considering the ionization losses, we can estimate that the maximum energy of accelerated electrons in RREA is about 60 MeV.

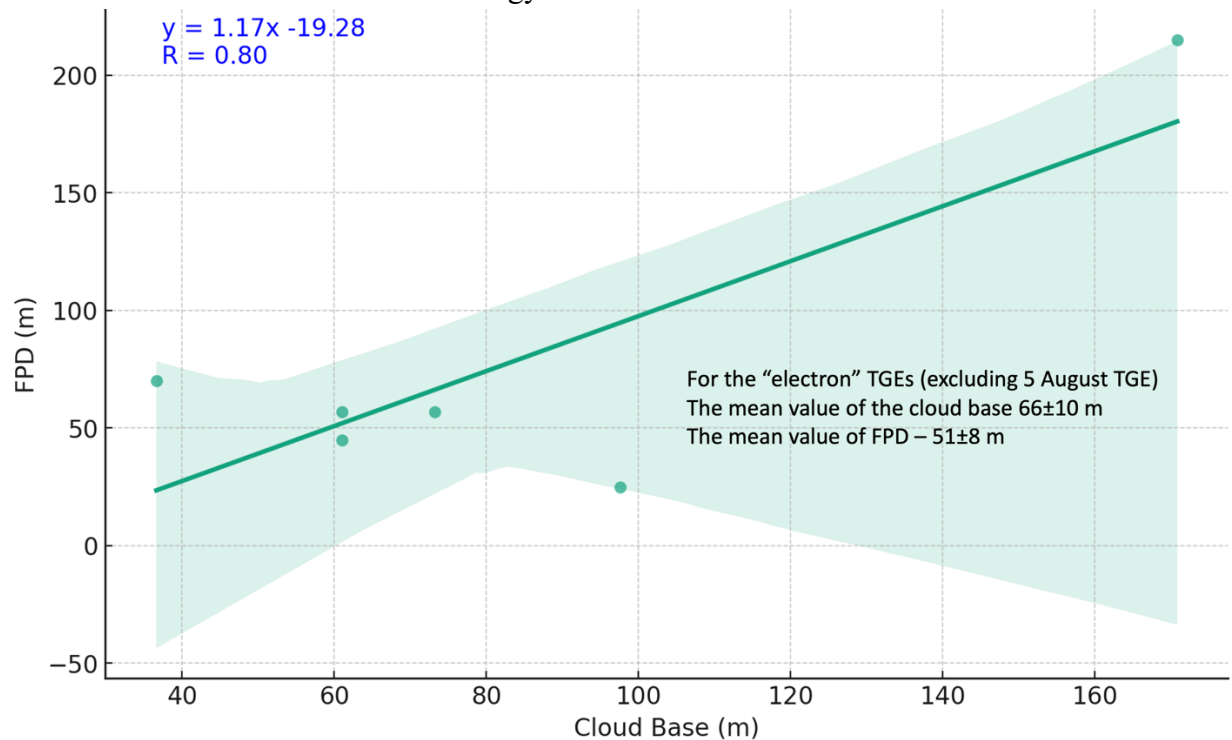

**Figure 8. Dependence of the distance to cloud base from the ground estimated by spread (equation 2) on free path distance estimated by equation (1)**

The low location of cloud base height, measured TGE intensities, and recovered energy spectra indicate that a strong electric field with a strength of at least 2.1 kV/cm can be present 50 m above ground. 2.1 kV/cm is the minimal strength of the electric field, which can sustain the RREA process at a height of 3250 m (Chilingarian et al., 2021a).

5. **Further analysis of TGE data**

We have developed an open-access database to facilitate the community's access to TGE data. In support of this initiative, we have compiled summary information and housed it within the Mendeley datasets, complete with detailed explanations (Chilingarian et al., 2024a). The Excel table illustrated in the accompanying images enumerates 56 TGE events, providing direct links to the associated graphical representations in the Cosmic Ray Division (CRD) database of the Yerevan Physics Institute (ADEI, Chilingaryan et al., 2008) with wide multivariate visualization and statistical analysis possibilities.

Our objective is to offer comprehensive multivariate data enabling researchers to examine potential correlations between charged and neutral particle fluxes, electric field strengths, lightning events, and meteorological parameters. This is made possible by integrating the ADEI visualization and correlation analysis platform, supplemented by the Excel files available in the Mendeley datasets.

Leveraging the advanced analytics powered by ChatGPT from (OpenAI 2024), we simplify the extraction of correlation insights from extensive Excel datasets, ensuring the process is user-friendly and efficient. Below, we present examples of further analysis of TGE data performed with ADEI and ChatGPT. The images that follow demonstrate the robust analytical capabilities that our platform supports.

For the August 5 TGE (Fig. 9), although the TGE flux enhancement was 115% (see Table 1), no significant electron flux was detected. The estimated cloud base height was 170 m and FPD 215 m; the electron spectrum attenuates below the recovery possibility. No coincidences of "1110" were registered for STAND3, and only 660 particles were counted for "1100" coincidence, which is the minimum count among the five largest TGEs of 2023. The electron energy release histogram, shown in the inset to Fig. 9, suggests their origin from the huge gamma-ray flux that travels from ≈200 m after going out of the accelerating electric field (Williams et al., 2023, Chilingarian et al., 2023). Nearly uniformly distributed 100 electrons in the energy release histogram possibly originated from the Compton scattering of an intense gamma-ray flux during ≈200 m travel in the dense air (Williams, 2024).

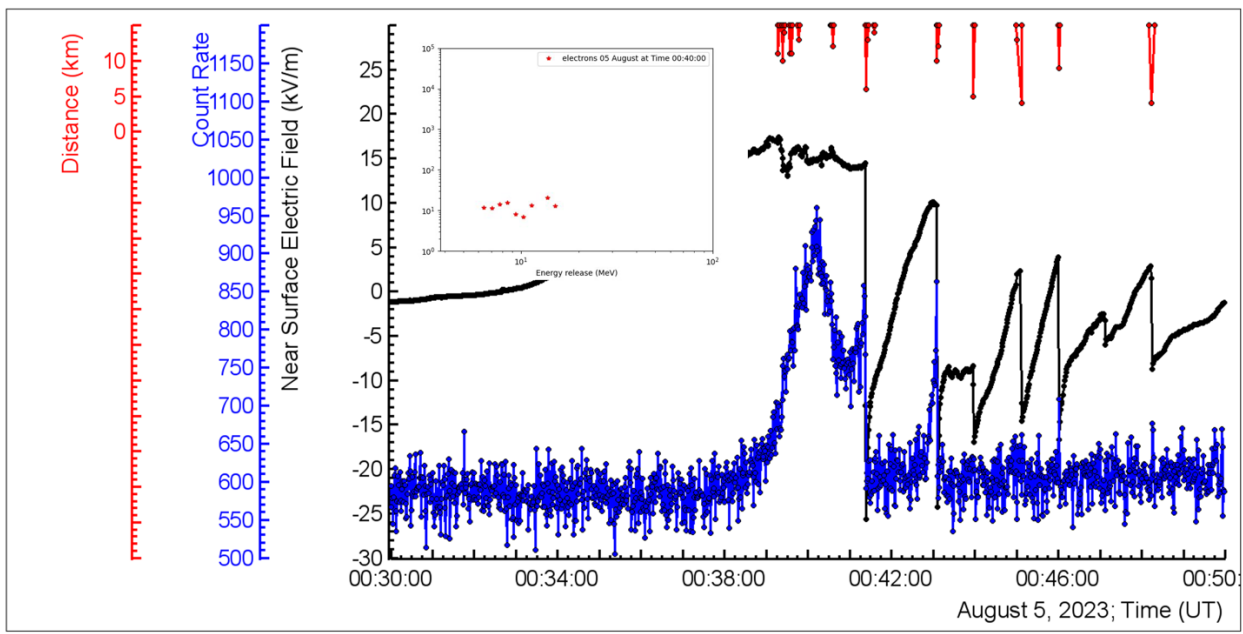

**Figure 9. The time series of STAND1's upper scintillator was abruptly terminated by a lightning flash (blue), disturbances of NSEF (black), and distances for the lightning flash (red). In the inset, we show the energy release histogram of TGE electrons.**

Figure 10 shows another remarkable event with a very long duration and enormous fluence. The TGE duration was 32 minutes, no lightning flash was detected, and TGE ended rather smoothly.

The fluence of the event was very large, 38 particles/cm² for particles with energies above 1 MeV.

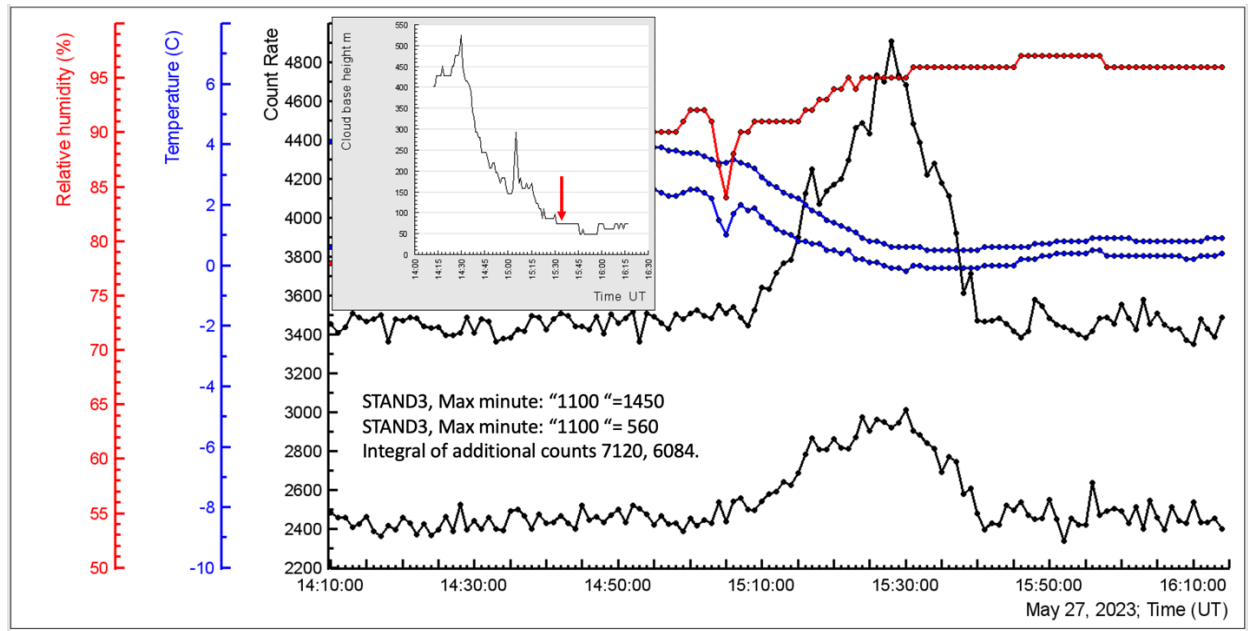

**Figure 10. 1-minute time series of STAND's 1100 (upper black) and 1110 (lower black) coincidences; blue- temperature and dew point; red – relative humidity. In the inset, the distance to the cloud base, by the red arrow, the distance at TGE maximum is denoted.**

The maximum energies of TGE gamma rays and electrons measured by the CUBE spectrometer above the roof of the experimental hall were 48 and 46 MeV, respectively (Fig. 11). The small difference in maximum energies (a sign of the low location of the atmospheric electric field) is confirmed by the large intensity of STAND's coincidences shown in Fig.10. The FPD calculated by Eq. 1 is ≈ 70 m, in good agreement with the cloud base height estimate (57 m). The accuracy of both estimates is smaller than 50 m (Chilingarian et al., 2021a, supplemented materials).

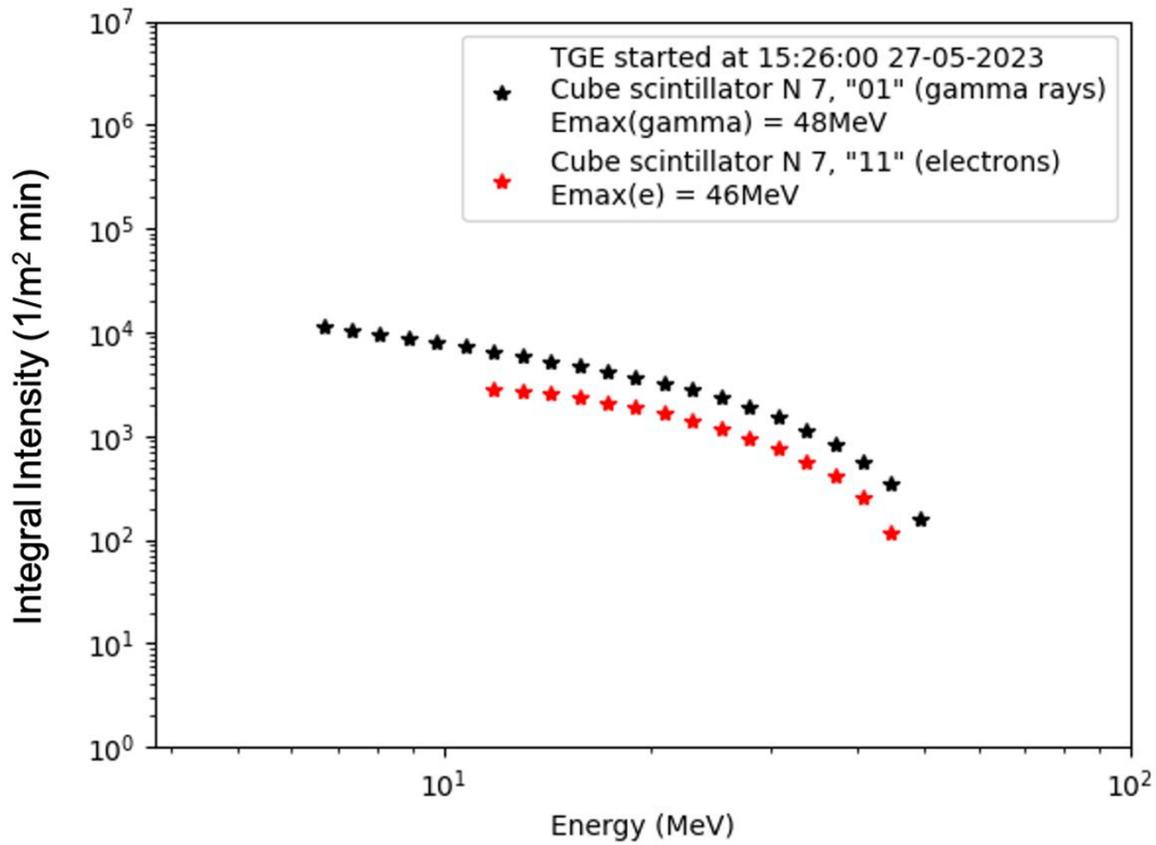

**Figure 11. The CUBE spectrometer recovered the integral energy spectra of gamma rays (|black) and electrons (red).**

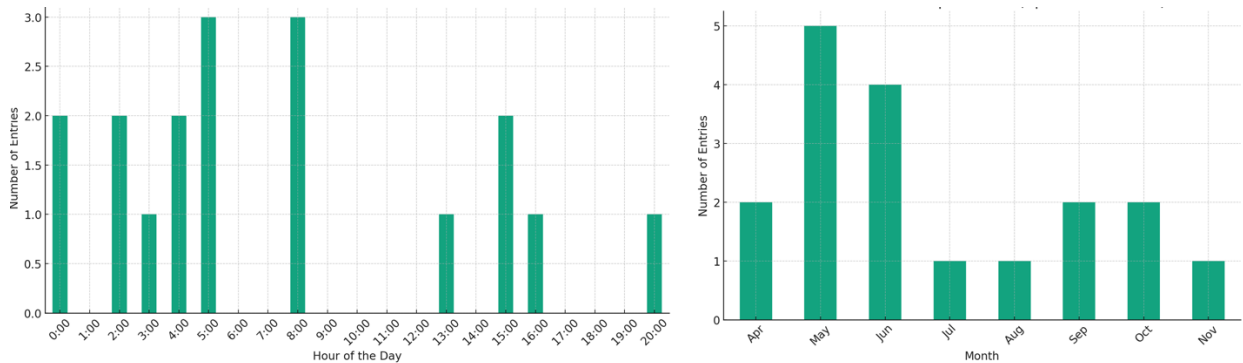

**Figure 12. Daily and monthly distributions of the TGEs registered in 2023 (largest TGEs with enhancement > 20% according to STANd3's "1000" coincidence.**

As we see in Figure 12, most (10 from 18) large TGE occurred at night-morning of the day. The monthly distribution outlines May and June as the most frequent months for TGE occurrence (9 from 18).

## 2023 TGEs terminated by flash and finishing smoothly: distance to flash

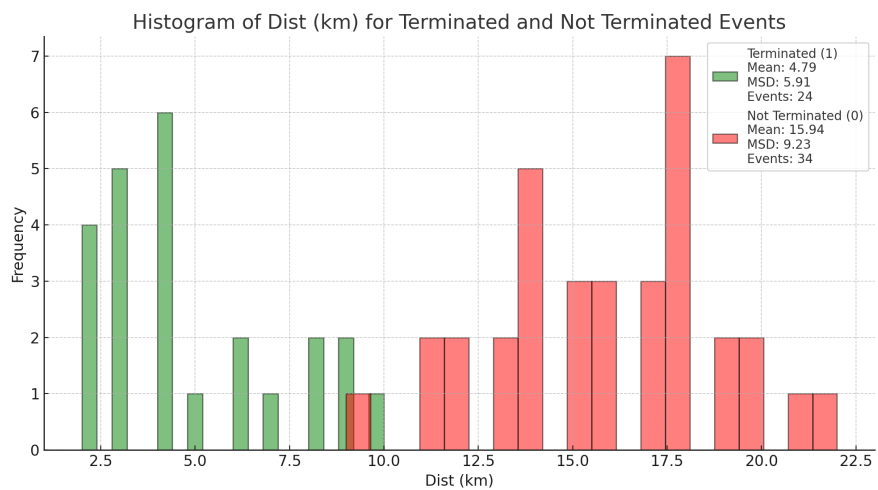

**Figure 13. Histogram of the distances to the lightning flash for 2023 TGEs, those that abruptly ended by flash (in green) and those that smoothly finished (in red).**

Figure 13 displays the distribution of TGE events based on the proximity to the nearest flash detected by BOLTEK's EFM 100 sensor. It illustrates that TGEs can only be terminated by flashes less than 10 km away (green). If no flashes were detected during TGEs, they smoothly ended after a few tens of minutes. From Figure 13, it can be inferred that TGEs serve as precursors to lightning flashes (Chilingarian et al., 2017).

### 6. Conclusions

- Based on the Thunderstorm ground enhancements (TGEs) recorded in 2023, we substantiate that extensive electric fields catalyze Relativistic Runaway Electron Avalanches (RREAs) in extensive regions within thunderclouds. These RREAs generate TGEs that cover multiple square kilometers on the Earth's surface.

- We have observed a seasonal and diurnal pattern in TGE occurrences, with a higher frequency during May and June and predominantly in the night-to-morning hours.

- Most intense TGE events above Aragats exhibited a tenfold increase in particle flux intensity, with TGE electron energies surging up to 60 MeV. Concurrently, the TGE fluence was measured at 38 particles per square centimeter.

- Our research indicates that atmospheric electric fields, associated with the TGEs with large electron content, can be as strong as 2 kV/cm and be present as low as 50 meters above the ground.

- Despite the chaotic nature of atmospheric electric fields, electron accelerators within thunderclouds demonstrated remarkable stability, maintaining a steady flux for durations of 0.5 to 2 minutes. The relative error of TGE flux at these minutes was lower than that associated with the ambient cosmic ray population, indicating a high level of stability of the electron accelerator. This finding indicates a level of organization within the atmospheric electric fields that was previously unappreciated.

- Interestingly, during TGEs, there was a marked suppression of lightning activity. Lightning flashes that terminated TGEs were typically within a 10 km radius. When lightning occurred farther than 10 km, the TGEs had a prolonged duration and concluded smoothly. These observations support the hypothesis that RREAs may be precursors to lightning flashes by developing ionization channels for lightning leaders. These observations contribute to our understanding of the elusive mechanisms behind lightning initiation.

- We introduce 2 empiric parameters, free passage distance, and cloud base height, to characterize the location of the vertical atmospheric electric field. These parameters are well correlated with the measured intensity of the TGE fluxes.

- Monitoring the maximum energy of the electron flux during thunderstorms with a simple spectrometer can be useful for alerting on extreme near-surface electric fields that can be dangerous during rocket launch and charging.

## ACKNOWLEDGMENTS

We sincerely thank the Aragats Space Environmental Center staff for their seamless operation of experimental facilities on Mount Aragats. S.B. and K.T. thank Hovsepyan G. for helping recover the energy spectra measured by the CUBE detector. A.C. thanks Earle Williams for the useful comments. This research effort has been made possible through the support of the Science Committee of the Republic of Armenia, Research Project No. 21AG-1C012.

**Data Availability Statement**

The data underpinning this study can be accessed in numerical and graphical formats through the multivariate visualization software platform ADEI, hosted on the Cosmic Ray Division (CRD) webpage of the Yerevan Physics Institute (ADEI, 2024).

38